Research Note

# ChatGPTest: opportunities and cautionary tales of utilizing AI for questionnaire pretesting

Francisco Olivos[1]

Department of Sociology and Social Policy, Lingnan University, Hong Kong SAR

Minhui Liu

Department of Sociology, The University of Hong Kong, Hong Kong SAR

Conflict of Interest
None.

Acknowledgements
This idea originated from the Sociological Research Methods course for undergraduate students taught by the corresponding author at Lingnan University, Hong Kong SAR. Thanks to the students who were willing to share part of their class assignment for this illustration and Tobias Kamelski for his comments.

Word count: 2381 and 2 Tables.

[1]Corrsponding author. E-mail: franciscoolivosrave@LN.edu.hk

Research Note

# ChatGPTest: opportunities and cautionary tales of utilizing AI for questionnaire pretesting

## Abstract

The rapid advancements in generative artificial intelligence have opened up new avenues for enhancing various aspects of research, including the design and evaluation of survey questionnaires. However, the recent pioneering applications have not considered questionnaire pretesting. This article explores the use of GPT models as a useful tool for pretesting survey questionnaires, particularly in the early stages of survey design. Illustrated with two applications, the article suggests incorporating GPT feedback as an additional stage before human pretesting, potentially reducing successive iterations. The article also emphasizes the indispensable role of researchers' judgment in interpreting and implementing AI-generated feedback.

Keywords: ChatGPT, survey design, questionnaire pretest, large language models.

## 1. Introduction

In survey methodology, various techniques like expert reviews, focus groups, cognitive interviews, and field pretests are employed for evaluating survey questions before data collection (Groves et al., 2009). Pretesting involves a small-scale trial of data collection procedures, aiming to assess them. It is a standard practice in both academic and applied research (Grimm 2010), and the output of the pretest is usually the feedback offered by interviewers on how to improve procedures and questions.

The rapid advancements in generative artificial intelligence (GAI) have opened up new avenues for enhancing various aspects of research, including the design and evaluation of survey questionnaires. AI technologies like large language models (LLMs) have demonstrated remarkable potential in generating human-like text, offering a promising approach to pretesting survey instruments. This article explores the use of GPT models as a tool for pretesting survey questionnaires. Illustrated with two applications, it suggests incorporating GPT feedback as an additional stage before human pretesting, potentially reducing successive iterations. However, the article emphasizes the indispensable role of researchers' judgment in implementing AI-generated feedback.

GPT is an LLM that utilizes advanced algorithms to generate texts that mimic the syntax, semantics, and grammar of human writing, which are approximated by statistical patterns learned from training data (for a technical review, see OpenAI 2023). Like most of the LLMs, GPT models predict the next word in a sequence based on the preceding words. Thus, when we provide a prompt asking to evaluate survey questions, a GPT will interpret the prompt and provide an answer accordingly based on its learned knowledge. It will retrieve the relevant patterns and associations from its trained data and generate a coherent and contextually relevant answer. We utilize OpenAI's GPT-4, which has exhibited human-level performance on different benchmarks, and outperformed other LLMs (Antaki et al. 2023; Nori et al. 2023; Katz et al. 2024).

## 2. Application to students' assignment

A good practice in prompt engineering is to provide the intent and context that describes the problem the prompt pattern attempts to solve (White et al. 2023). We may indicate that this is a questionnaire pretest and want feedback on our question. However, additional information in the prompt could lead to different feedback from the GPT which may be considered during the interaction. For instance, in addition to the particular task as part of the pretest, we can specify other key ideas such as (1) the aim of the research, (2) the survey mode, and (3) the target population of the study.

As part of an introductory course to social research at Lingnan University, students conducted a small-scale questionnaire-based study about a topic of their interest. In Table 1, we displayed different AI-generated suggestions provided by ChatGPT 4 to one of their questions[1]. The question proposed was: "*How frequently do you engage in activities within natural environments (such as parks and gardens) outside your university campus?*" With answers in three possible categories: (1) Never, (2) 1-2 days a week, (3) 3-4 days a week. Prompts with increasing levels of complexity were utilized in independent tasks. We also indicated that ChatGPT does not need to provide a revised question, which is included in the next application.

[Table 1 about here]

The answers to the different prompts show commonalities and differences. First, we only indicated the purpose of seeking feedback to improve the question and categories as part of a questionnaire pretest. ChatGPT was successful in identifying major issues with this question. According to the feedback, the quality of the question can increase if the terms "activity" and "outside of the university" are further specified. Moreover, the most evident problem is the completeness of the response categories because respondents can engage more than 4 days a week, and this option is not included. It also suggests including an intermediate category between weekly and never, which could add more detail to the data collected. The temporal frame is another possible improvement suggested by ChatGPT. It can be solved by adding a "typical week" or another time of the year. However, there are two aspects that highlight the importance of researchers' judgment. Regarding the timeframe, one of the ideal solutions is to ask about the last 7 days, which ChatGPT has not given a solution. Additionally, ChatGPT is right on the need for a neutral tone and avoiding leading responses in questionnaire design; nonetheless the original question has no major issues in this regard. Therefore, researchers could be puzzled by this feedback and introduce unnecessary changes.

---

[1] Students consented the use of these question for this article.

Second, the prompt was modified by adding the aim of the study in one sentence. The feedback has commonalities with the previous one, as well as certain differences. Regarding commonalities, ChatGPT also identified the need to expand the response options and specify the temporal frame. In contrast, the new feedback omits the ambiguity of activities and suggests specifying a definition of the natural environment with examples when it has already been included in the original question. Moreover, a non-existent overlap between response categories is suggested. Therefore, although the feedback is useful, the researcher may critically assess the suggestions.

The population was specified in the third prompt. Again, ChatGPT recommended specifying the type of activities, the response options, and the timeframe. There is only a variation in comparison to the previous feedback regarding the location, where only an emphasis is recommended but not specification as in the first prompt.

In the last task, we explicitly include the ten principles of good question writing proposed by Neuman (2010). The principles can be discussed and adjusted according to the research purposes, but they add more specific instructions to the prompt and clearer guidelines to ChatGPT. The main change is that the revised question and categories appear up to front, followed by the feedback and rationale for changes. Both activities and natural environment are clarified and specified by replacing activities with visits, and including parks and gardens as part of the sentence rather than in parentheses. Similarly to the previous prompts, ChatGPT recommends improving the response categories. Moreover, only with this prompt does the answer reference the language's complexity as a rationale to replace "engaging in activities within" with "visit."

Overall, the simplest prompt already provided insights into the most evident aspects to be improved in this question. Higher levels of complexity lead to different suggestions, but the key elements of the initial feedback are maintained.

### 3. Comparison to expert judgment

The second application of AI for questionnaire pretesting utilizes the text features pointed out by Lenzner (2012) that affect respondents' processing efforts. We prompted ChatGPT to provide feedback on the ambiguity and comprehensibility of questions used by Lenzner (2012) to illustrate difficulties in comprehension. Additionally, we explicitly asked ChatGPT to role play as a respondent from the general population to provide comments and as an expert to suggest a revised version of the question[2]. We compare Lenzner's suggested improvements

---

[2] More advanced applications of GPT could include different profiles to pretest the instrument with specific

with the AI-powered questions. The prompt utilized was:

> *"We want to do a questionnaire pretest with you. Imagine you are a participant from the general population and evaluate the unambiguity and comprehensibility of the question. Less cognitive effort and ambiguity may reduce comprehension difficulties and response error. Then, provide an improved version at the end of your comment as an expert, but only if it can be improved. This is the question: [Question]"*

The results are reported in Table 2.

[Table 2 about here]

The AI-generated feedback identified the comprehension difficulty generated by the low-frequency term "somatic" in the original question (#1). However, there are two additional changes that should be highlighted. First, ChatGPT included examples of physical pain to make it even clearer, which is not contained in Lenzner's improved version. Second, ChatGPT replaced "4 weeks" with "past month," which can increase ambiguity. Thus, this application highlights the contribution of ChatGPT to improving questions but also requires the critical assessment of the research to evaluate the appropriateness of the proposed timeframe, an aspect in which the original question and Lenzner's proposal are superior.

Moreover, ChatGPT identified the need to replace vague relative terms (#2). In the question in Table 2, ChatGPT indicates that the term "recently" is ambiguous and suggests a timeframe, which is also done by Lenzner. Nevertheless, the AI-generated feedback pointed out that "doctor" could be ambiguous and replaced "seen" by visit. These changes require a contrast with human subjects or consideration of usage in the particular context of the study. Nevertheless, it points out additional aspects to be considered in the following stages of the questionnaire design. Similarly, ChatGPT suggests to disambiguating cultural events in the third question (#3).

In the fourth case, ChatGPT successfully suggested to reduce the level of complexity of the question (#4). Similarly to Lenzner's proposal, the AI-proposal is simpler and conveys the same meaning with even fewer words. In this case, this is a good example of how GPT can, in certain occasions, outperform a human expert.

Furthermore, ChatGPT could simplify complex logical structures (#5). However, AI-generated question differs from the question proposed by Lenzner. ChatGPT keeps a high number of coordinated phrases in a single-sentence question. In contrast, Lenzner's question is separated into three sentences. Again, researchers' judgment may play a substantial role in discerning which suggestions to incorporate. In the case of Lenzner's question, the proposal removes "at least somewhat oppose," which changes what is being measured by this question.

populations.

In contrast, ChatGPT did not remove this option.

According to Lenzner (2012), another text feature that affects comprehension is the low syntactic redundancy (#6). The lack of redundancy makes it difficult to understand questions, and it is increased by denominalizing nominalizations. ChatGPT fails to recognize the difficulty generated by the nominalization "security" (i.e., the use of a verb as a noun). Nevertheless, AI suggested a change in the term "important" due to potential ambiguity. Additionally, ChatGPT added "to what extent" before "do you agree or disagree." Although the rationale is not clear, the phrase provides the possibility of nuanced responses.

Finally, as shown in the last row of Table 2 (#7), ChatGPT and Lenzner coincide in the difficulty posited by the need to connect "making mistakes" to "convicting an innocent person or to let a guilty person go free." However, they offered two different wordings as solutions to this issue. While Lenzner added "veredict," ChatGPT referred to problematic outcomes for a justice system. Additionally, ChatGPT pointed out the possible ambiguity of "worse," explaining that for some people, it could be morally wrong, but for others, it can mean negative practical consequences.

## 4. Conclusions

In this article, we highlighted the benefits of using GPT for AI-generated questionnaire pretesting, demonstrating its usefulness in early survey design. However, it underscores the substantial role of researchers in critically identifying solutions that GPT misses or unnecessary changes it proposes.

Although AI-generated feedback enhances question quality, human pretesting is still essential for several reasons. First, AI can be algorithmically biased (Fazelpour and Danks 2021). For instance, GPTs are trained with large amounts of data from the Internet, which can exclude language usage of people underrepresented in the training dataset. Second, GPT not only provides feedback on the improvement of questions but also suggests new questions and probes that need to be further tested. Third, while GPT outperforms humans at the bottom of the distribution of cognitive performance (Katz et al. 2024; Martínez 2024), it is less clear whether increasingly complex LLMs can replicate cognitive tasks of cognitively bounded individuals who are also subjects of surveys.

Overall, the applications of AI for questionnaire pretesting have significant potential, offering tailored feedback based on diverse population profiles. For instance, researchers can simulate different population profiles following a theoretical sample to obtain feedback tailored to specific demographics. Researchers can have more detail suggestions on how to improve survey for heterogeneous populations. Furthermore, GPT can serve as a valuable tool in teaching and learning survey design, providing students with in-class feedback on their questions or opportunities to evaluate AI-generated suggestions.

Table 1. Sequential prompting

| # | Prompt type | Prompt | Feedback |
|---|---|---|---|
| 1 | Only task | This is part of a questionnaire pretest. Provide feedback to improve the following question and its categories. You do not need to give me a revised version, only the feedback: [Question] | 1. **Clarify the Definition of "Activities":** Specify what types of activities you're referring to (e.g., walking, exercising, relaxing, studying). This helps respondents better understand what you're asking.<br>2. **Include All Relevant Locations:** Since the question specifies "outside your university campus," it might be worth clarifying whether this includes areas immediately surrounding the campus or only more distant locales.<br>3. **Expand the Frequency Options:** The current frequency categories may miss out on capturing data from those who engage in such activities less than once a week but more than never. Additionally, consider including an option for those who do so more frequently than 4 days a week.<br>4. **Consider Different Temporal Patterns:** People's engagement with natural environments might vary by season or school term. Consider asking about frequency in a typical week, perhaps specifying a particular time of year if relevant.<br>5. **Use a Clear and Neutral Tone:** Ensure the question is phrased in a way that is neutral and straightforward to avoid leading responses. |
| 2 | Task and research question | This is part of a questionnaire pretest. Provide feedback to improve the following question and its categories. The study aims to understand the relationship between the natural environment and educational performance. You do not need to give me a revised version, only the feedback: [Question] | 1. **Clarify the Definition of "Natural Environments":** Make sure that respondents have a clear understanding of what constitutes "natural environments." Specify examples or characteristics, e.g., parks, forests, gardens, lakesides.<br>2. **Expand Response Options:** The current options might not capture all relevant frequencies, especially for those who engage with natural environments less frequently than weekly but more than "never." Additionally, consider covering those who might visit these environments daily or almost daily.<br>3. **Consistency in Option Ranges:** Ensure that the response categories are mutually exclusive and collectively exhaustive. There's a slight overlap in the current categories as someone who goes 2 days might select either (2) or (3).<br>4. **Add a Time Frame:** Specify a time frame for better accuracy, such as "in a typical week" or "during the past month." |
| 3 | Task, research question, and population | This is part of a questionnaire pretest. Provide feedback to improve the following question and its categories. The study aims to understand the relationship between the natural environment and educational performance. The population of the study is university students. You do not need to give me a revised version, only the feedback: [Question] | 1. **Clarify the Type of Activities**: Specify what types of activities you are referring to, as "engaging in activities" can be interpreted differently by different respondents. It might be helpful to specify whether you mean studying, relaxing, exercising, etc.<br>2. **Extend the Range of Response Options**: The current categories may not capture the full range of possible responses. It might be beneficial to include options for those who engage in activities less frequently than once a week but more than "never."<br>3. **Add More Frequency Options**: The gap between the categories could miss important variations in engagement frequency. Consider adding more categories to capture finer distinctions. |

| | | | 4. **Specify the Time Frame**: It's unclear over what period the frequency should be assessed. Adding a time frame, such as "per week" or "per month," at the beginning of the question could clarify this. |
| --- | --- | --- | --- |
| | | | 5. **Consider the Relevance of Location**: Since the focus is on activities outside the university campus, reaffirm this in the question to avoid confusion about where the activities should take place. |
| 4 | Task, research question, population, and principles of good question writing | This is part of a questionnaire pretest. Provide feedback to improve the following question and its categories. The study aims to understand the relationship between the natural environment and educational performance. The population of the study is university students. In addition to general principles, you should consider taking into account the following: avoid jargon, and abbreviations; avoid ambiguity, confusion, and vagueness; avoid emotional language and prestige bias; avoid double-barreled questions; avoid leading questions; avoid asking questions that are beyond respondents' capabilities; avoid false premises; avoid asking about future intentions; avoid double negatives; avoid overlapping or unbalanced response categories. You do not need to give me a revised version, only the feedback: [Question] | 1. **Rephrase the question to make it more clear and concise**: "How often do you participate in activities in natural environments (e.g., parks, gardens) outside your university campus?" <br> 2. **Revise the categories to include more options:** The current answer categories are not exhaustive and may not accurately capture the range of responses. Consider revising the categories to include more options, such as: - Never - Less than once a month - 1-3 times a month - Once a week - 2-3 times a week - 4-5 times a week – Daily <br> 3. **Clarify the definitions of activities:** Provide examples of the activities in natural environments to give respondents a clearer idea of what types of activities you are referring to. For example: "Examples of activities include walking, jogging, picnicking, or participating in sports." |

Table 2.

| # | Text feature | Original question | AI-generated feedback | AI proposal | Lenzner's proposal |
|---|---|---|---|---|---|
| 1 | Low-frequency words | During the last 4 weeks, how often did you suffer from *somatic* pain? | As a participant, the question "During the last 4 weeks, how often did you suffer from somatic pain?" is clear and straightforward. However, "somatic pain" might not be a term everyone is familiar with, potentially leading to some confusion. Also, "how often" might be interpreted differently by different individuals, leading to varied responses. Overall, it's a decent question, but it could be clearer. | Over the past month, how frequently have you experienced physical pain, such as headaches, muscle aches, or joint pain? | During the last 4 weeks, how often did you suffer from physical pain? |
| 2 | Vague or imprecise relative terms | Have you *recently* seen a doctor? If yes, please provide the number of visits you paid to the doctor. | As a participant, I understand the question clearly. It asks whether I've visited a doctor recently and if so, how many times. However, the term "recently" might be ambiguous as it can vary from person to person. Some might consider "recently" as within the past month, while others might think of it as within the past year. Additionally, it doesn't specify what type of doctor visits it's referring to (e.g., general check-up, specialist appointment, emergency visit). Overall, the question could be improved for better clarity. | Have you visited any type of healthcare provider for medical reasons within the last three months? If yes, please specify the total number of visits you have made | Have you seen a doctor during the last 4 weeks? If yes, please provide the number of visits you paid to the doctor. |
| 3 | Vague or ambiguous noun phrases | In your free time, how often do you attend *cultural events*? | As a participant, I understand the question, but it could be clearer. "Cultural events" is quite broad, so I might wonder what exactly it includes. Does it mean concerts, art exhibitions, theater performances, or something else? Also, "how often" might be interpreted differently by different people. Some might think in terms of days per week, others in terms of times per month or year. Adding more specificity could make it easier to answer accurately. | How frequently do you participate in cultural events such as concerts, art exhibitions, theater performances, or similar activities in your leisure time? | In your free time, how often do you go to the theather? |
| 4 | Complex syntax | How likely is it that if a law was considered by parliament that you considered to be unjust or harmful, you, acting alone or together with others, would *try to do* something against it? | As a participant from the general population, I find the question somewhat ambiguous and difficult to understand due to the long sentence structure and multiple conditions. It may be challenging to comprehend the question without re-reading it a few times. | How likely are you to take action, either individually or with others, if you believe a law proposed by parliament is unjust or harmful? | How likely is that you, acting alone of together with others, would try to do something against a law that was considered by parliament and that you believed to be unjust or harmful? |
| 5 | Complex logical structures | There are many ways people *or* organizations can | As a participant, I find the question to be clear and easy to understand. It asks whether organizing public meetings to | Do you believe it should be permissible for individuals or | There are many ways people or organizations can |
| | | protest against a government action *or* a government plan they strongly *or* at least somewhat oppose. In this regard, do you think the following should be allowed? Organizing public meetings to protest against the government. | protest against the government should be allowed, which is straightforward and doesn't leave much room for confusion. However, there might be some room for improvement to make it even clearer. Here's a revised version. [Proposal] This version maintains clarity while specifying the purpose of the meetings as a form of protest. | groups to organize public meetings as a form of protest against government actions or plans they strongly or somewhat oppose? | protest against a government they strongly oppose. In this regard, do you think the following should be allowed? Organizing public meetings to protest against the government. |
| 6 | Low syntactic redundancy | Do you agree or disagree with the following statement? Trade unions are important for the job *security* of employees. | As a participant from the general population, I find this question to be quite clear and unambiguous. It is asking for my opinion on whether trade unions have a significant role in ensuring job security for employees. There is no confusion or cognitive effort required to understand the question, and it appears to be a straightforward agree or disagree response.<br><br>However, if I were to suggest a minor improvement to the question, I would recommend adding a scale for the level of agreement or disagreement to allow for more nuanced responses | To what extent do you agree or disagree with the following statement? Trade unions play a significant role in ensuring the job security of employees.<br>To what extent do you agree or disagree with the following statement? Trade unions are important for the job security of employees.<br><br>- Strongly agree<br>- Somewhat agree<br>- Neither agree nor disagree<br>- Somewhat disagree<br>- Strongly disagree | Do you agree or disagree with the following statement? Trade unions are important to secure the jobs of employees. |
| 7 | Bridging inferences | All systems of justice *make mistakes*. What do you think is worse, to convict an innocent person or to let a guilty person go free? | As a participant, I understand the question, but it feels a bit broad and could be interpreted differently by different people. It's clear that it's asking about the relative importance of convicting innocent people versus letting guilty ones go free in the context of justice systems, but the terms "worse" and "mistakes" might be open to interpretation. Some might interpret "worse" in terms of moral implications, while others might think in terms of practical consequences. Additionally, "mistakes" might not fully capture the complexity of justice system failures. | In your opinion, which outcome do you consider more problematic for a justice system: convicting an innocent person or letting a guilty person go free? | All systems of justice make wrong verdicts. What do you think it is worse, to convict an innocent person or to let a guilty person to go free? |